\documentclass[10pt,twocolumn,a4paper]{aastex63}
\usepackage{graphicx}	
\usepackage{amsmath}	
\usepackage{amssymb}	
\usepackage{color}
\usepackage{array}
\usepackage{lineno}
\usepackage{bookmark}
\usepackage{subfigure}

\shorttitle{Dark matter annihilation in gamma-ray burst afterglows}
\shortauthors{Huang et al.}

\begin{document}
\title{Contribution of dark matter annihilation to gamma-ray burst afterglows near massive galaxy centers}

\correspondingauthor{Tong Liu}
\email{tongliu@xmu.edu.cn}

\author{Bao-Quan Huang}
\affiliation{Department of Astronomy, Xiamen University, Xiamen, Fujian 361005, China}
\author{\href{https://orcid.org/0000-0001-8678-6291}{Tong Liu}}
\affiliation{Department of Astronomy, Xiamen University, Xiamen, Fujian 361005, China}
\author{\href{https://orcid.org/0000-0002-6654-3716}{Feng Huang}}
\affiliation{Department of Astronomy, Xiamen University, Xiamen, Fujian 361005, China}
\author{\href{https://orcid.org/0000-0003-1474-293X}{Da-Bin Lin}}
\affiliation{Laboratory for Relativistic Astrophysics and Department of Physics, Guangxi University, Nanning 530004, China}
\author{\href{https://orcid.org/0000-0002-9725-2524}{Bing Zhang}}
\affiliation{Department of Physics and Astronomy, University of Nevada, Las Vegas, Las Vegas, NV 89154, USA}

\begin{abstract}
Gamma-ray bursts (GRBs) are believed to be powered by ultrarelativistic jets. If these jets encounter and accelerate excess electrons and positrons produced by particle dark matter (DM) annihilation, the observed electromagnetic radiation would be enhanced. In this paper, we study GRB afterglow emission with the presence of abundant DM under the weakly interacting massive particle annihilation conditions. We calculate the light curves and spectra of the GRB afterglows with different parameters, i.e., DM density, particle DM mass, annihilation channel, and electron density of the interstellar medium. We find that the effect of DM may become noticeable in the afterglow spectra if the circumburst  has a low electron number density ($n \lesssim 0.1~\rm cm^{-3}$) and if the DM has a high number density ($\rho_\chi \gtrsim 10^3~\rm GeV~cm^{-3}$). According to the standard galaxy DM density profile, GRB afterglows with DM contribution might occur at distances of several to tens of parsecs from the centers of massive galaxies.
\end{abstract}


\section{Introduction}

The existence of dark matter (DM) is strongly supported by convincing evidence in cosmology and astrophysics \citep[e.g.,][]{Bertone2005}. Numerous astronomical processes and evolutions are partly or predominantly affected by DM. However, the nature of DM remains a mystery, and its direct detection has yet to be achieved. The mainstream DM model invokes DM particles, the most promising candidate being the weakly interacting massive particles \citep[WIMP, see e.g.,][]{Jungman1996,Bertone2005,Bergstrom2012}. The WIMP model can be tested by detecting potential signals from WIMP annihilations, which can ultimately produce Standard Model particles, such as neutrinos, photons, electrons, and positrons. Searching for gamma-ray emissions produced by DM annihilation is one method to indirectly detect DM; for this task, nearby galaxy centers and dwarf spheroidal galaxies are appealing targets \citep[e.g.,][]{Ackermann2015,Abdallah2018,Johnson2019}. Moreover, the radio detection of synchrotron radiation induced by electrons and positrons produced by DM annihilation in galaxies is considered a promising method to constrain the particle nature of DM \citep[e.g.,][]{Storm2013,Egorov2013}. However, such emissions are generally weak, and possible detections are limited to galaxies in the local universe.

If significant signals can be powered by electrons produced by DM annihilation, an accelerator should be established in a high-density DM halo. In addition, the number density of system-provided electrons should be comparable to that of DM electrons (DMEs). The jets generated from gamma-ray bursts (GRBs) near galaxy centers are ideal accelerators.

GRBs are the most luminous explosions in the universe. The ultrarelativistic jets in the line of sight launched by newly born magnetars or hyperaccreting black holes can trigger observable GRBs \citep[see the reviews by][]{Meszaros2006,Zhang2007,Zhang2018book,Liu2017}. The prompt gamma-ray emissions and multiband afterglows of GRBs generally originate from the internal and external shock phases, respectively \citep[e.g.,][]{Meszaros1993,Rees1994,Meszaros1997,Sari1998,Piran2004}. Most GRB afterglows can be explained by synchrotron and synchrotron self-Compton (SSC) processes under the conditions of external shocks sweeping the interstellar medium (ISM, or interstellar wind) and accelerating electrons. This explanation has been thoroughly verified by multi-wavelength observations of GRB 190114C \citep{MAGIC2019,Wang2019,Fraija2019}. In the prompt emission phase, the number of DMEs is much less than that of electrons emitted from the central engine, so the electromagnetic radiation of DMEs might be observable only in the afterglow phase.

In this paper, the lightest supersymmetric particle in the WIMP model, namely, the neutralino, is chosen; this particle has four different annihilation channels, i.e., $w^+w^-$, $b\bar{b}$, $\mu^+\mu^-$, and $\tau^+\tau^-$. We then study the acceleration and radiation mechanisms of DMEs in external shocks and subsequently predict and analyze the DM contributions to the light curves and spectra of GRB afterglows. In Section 2, we provide a detailed description of the method employed to calculate the light curves and spectra of GRB afterglows, including the effects of DM. In Section 3, the results using different DM parameters are shown. In Section 4, we analyze the locations of GRBs in massive galaxies associated with the effects of DM. The conclusions and discussion are presented in Section 5.

\section{Method}
\subsection{Dynamical evolution of the external forward shock}

As the GRB ejecta launched from the central engine interact with the circumburst medium, a relativistic shock is generated which  propagates through the medium. We adopt an approximate dynamical evolution model to discuss the evolution of the external forward shock \citep{Huang1999,Huang2000}
\begin{equation}
\frac{d \Gamma}{dm}=-\frac{\Gamma^{2}-1}{M_{\rm ej}+\epsilon m + 2 (1- \epsilon ) \Gamma m},
\end{equation}
where $\Gamma$, $M_{\rm ej}$, $m$, $\epsilon$ are the Lorentz factor, the ejecta mass, the swept mass from the external medium, and the radiation efficiency of the external shock, respectively. The problem can be solved by introducing two more differential equations
\begin{equation}
\frac{dm}{dR}=2 \pi R^{2}(1 - \cos\theta_{\rm j}) n m_{\rm p}
\end{equation}
and
\begin{equation}
\frac{dR}{dt} = \beta c \Gamma (\Gamma + \sqrt{\Gamma^2 - 1}),
\end{equation}
where $\theta_{\rm j}$ is the half opening angle of the ejecta, $n$ is the number density of the circumburst medium, $t$ is the time measured in the observer frame, $m_{\rm p}$ is the proton mass, and $\beta =\sqrt{1-1/\Gamma^2}$ is the ejecta velocity. Generally, the circumburst medium can be classified into two cases: an interstellar medium (ISM) and a stellar wind. The mass density of the circumburst medium is a constant in the former case and declines with $\sim R^{-2}$ in the latter case. In this work, we consider only the ISM case. We also neglect the evolution of $\theta_{\rm j}$ since numerical simulations showed that sideways expansion is not important \citep{ZhangW2009}.

\subsection{Electron distribution}

With the contribution of DMEs considered, the evolution of the shock-accelerated electrons can be expressed as a function of the radius $R$, i.e.,
\begin{equation}
\frac{\partial }{{\partial R}}\left(\frac{{dN_{\rm{e}}^\prime }}{{d\gamma _{\rm{e}}^\prime }}\right) + \frac{\partial }{{\partial \gamma _{\rm{e}}^\prime }}\left(\dot \gamma _{\rm{e}}^\prime \frac{{dt'}}{{dR}}\frac{{dN_{\rm{e}}^\prime }}{{d\gamma _{\rm{e}}^\prime }}\right) = {{\hat{Q}'}_{{\rm{ISM}}}} + {{\hat{Q}'}_{{\rm{DM}}}},
\end{equation}
where $dN_{\rm{e}}^\prime /d\gamma _{\rm{e}}^\prime $ is the instantaneous electron energy spectrum, $\gamma^\prime_{\rm e}$ is the Lorentz factor of the shock-accelerated electrons in the comoving frame of the shock, $dt'/dR$ = $1/\Gamma c$ (with $t'$ being the time in the shock comoving frame and $\Gamma$ being the Lorentz factor of the external forward shock), $\dot \gamma _{\rm{e}}^\prime $ is the cooling rate of electrons with the Lorentz factor $\gamma^\prime_{\rm e}$, and ${\hat{Q}'_{{\rm{ISM}}}}dR$ and ${\hat{Q}'_{{\rm{DM}}}}dR$ represent the injection of electrons from the ISM and DM into the shock during its propagation from $R$ to $R+dR$.

Here, $\hat{Q}'_{\rm ISM}=\bar{K} \gamma_{\rm e}^{\prime-p}$ with $\bar{K}\approx 4\pi(p-1)R^2n_{\rm}\gamma_{\rm e,min}'^{p-1}$ is adopted to describe the injection behavior of newly shocked circumburst medium electrons (CMEs) in the ISM, where $p$ ($>$ 2) is the power law index, $n$ is the number density of the circumburst medium, and $\gamma^\prime_{\rm e,min}\leq\gamma^\prime_{\rm e}\leq\gamma^\prime_{\rm e,max}$ is adopted for $\gamma^\prime_{\rm e}$.

The shocked electrons and the magnetic fields share the fractions $\epsilon_{\rm e}$ and $\epsilon_{B}$ of the thermal energy density in the forward shock downstream. Since DMEs are involved in our work, the minimum Lorentz factor of the shock-accelerated electrons can be expressed as $\gamma_{\rm e,min}'=\epsilon_{\rm e}\eta(\Gamma-1)(p-2) m_{\rm p}/(p-1)m_{\rm e}$, where $\eta$ ($\le$ 1) is the ratio of the number density of CMEs to the number density of all electrons, including both CMEs and DMEs, and $m_{\rm e}$ denotes the electron masses. The maximum Lorentz factor of electrons is $\gamma^\prime_{\rm e,max}=\sqrt{{9m_{\rm e}^{2}c^{4}}/{8B'e^3}}$, with $B^\prime=\sqrt{32 \pi \Gamma (\Gamma-1) n m_{\rm p} \epsilon_{B}c^2}$ being the magnetic field behind the shock \citep[e.g.,][]{Kumar2012}.

The evolution of the DME fluid is described by the following diffusion-loss equation \citep[neglecting re-acceleration and advection effects, see e.g.,][]{Colafrancesco2006,Borriello2009}, i.e.,
\begin{eqnarray}
 \frac{\partial}{\partial t}\frac{dn_{\rm e,\chi}}{d\gamma_{\rm e,\chi}}=&&\vec{\nabla}\cdot \bigg[K(\gamma_{\rm e,\chi},\vec{r})\vec{\nabla}
 \frac{dn_{\rm e,\chi}}{d\gamma_{\rm e,\chi}}\bigg]+
 \nonumber\\&&\frac{\partial}{\partial \gamma_{\rm e,\chi}}\bigg[b(\gamma_{\rm e,\chi},\vec{r})\frac{dn_{\rm e,\chi}}{d\gamma_{\rm e,\chi}}\bigg]+Q(\gamma_{\rm e,\chi},\vec{r}),
\end{eqnarray}
where ${dn_{\rm e,\chi}}/{d\gamma_{\rm e,\chi}}$ is the DME equilibrium spectrum, $K(\gamma_{\rm e,\chi},\vec{r})$ is the diffusion coefficient, and $b(\gamma_{\rm e,\chi},\vec{r})$ stands for the energy loss rate. Since GRBs are stellar-scale events, we reasonably assume that the values of the DM densities in these regions are constants. Then the gradients of the DME densities should be considered as 0, so the first term on the r.h.s of the above equation can be neglected. Thus, the steady-state energy spectrum for the DMEs in the interstellar medium can be expressed as \citep[e.g.,][]{Colafrancesco2006,Borriello2009}
\begin{equation}
\frac{dn_{\rm e,\chi}(\gamma_{\rm e,\chi})}{d\gamma_{\rm e,\chi}}=\frac{1}{b(\gamma_{\rm e,\chi},\vec{r})}\int\nolimits_{\gamma_{\rm e,\chi}}^{\gamma_{\rm e,\chi,max}}Q({\zeta,\vec{r}})d\zeta,
\end{equation}
where $\gamma_{\rm e,\chi,max}=M_{\chi}/m_{\rm e}$ is the maximum Lorentz factor of the DMEs, $M_{\rm \chi}$ is the DM mass, and $Q(\gamma_{\rm e,\chi},\vec{r})$ is the source item, which can be expressed as
\begin{equation}
Q(\gamma_{\rm e,\chi},r)=\frac{1}{2}\bigg(\frac{\rho_{\chi}(r)}{M_{\rm \chi}}\bigg)^2\langle\sigma_{A} \upsilon\rangle \frac{dN_{\rm inj}}{d\gamma_{\rm e,\chi}},
\end{equation}
where $\rho_{\chi}(r)$ is the DM density profile, $\langle\sigma_A \upsilon \rangle$ is the annihilation cross section that has a typical value ($3.0\times 10^{-26} \,{\rm cm^3 s^{-1}}$ is used in this work), and ${dN_{\rm inj}}/{d\gamma_{\rm e,\chi}}$ is the DME injection spectrum, which can be obtained by the Dark SUSY package \citep[e.g.,][]{Gondolo2004,Bringmann2018}. In addition, the energy loss term, $b(\gamma_{\rm e,\chi},\vec{r})$, involving the energy loss of synchrotron radiation and the inverse Compton scattering of the cosmic microwave background (CMB) and starlight photons, which are the faster energy loss processes for driving the equilibrium of DMEs, can be expressed by
 \begin{equation}
 \begin{split}
  b(\gamma_{\rm e,\chi},\vec{r})&=b_{\rm IC}(\gamma_{\rm e,\chi})+b_{\rm Synch.}(\gamma_{\rm e,\chi},\vec{r}) \\
                                &=b_{\rm IC}^{0} u_{\rm CMB} E_{\rm e,\chi}^2 + b_{\rm IC}^{0} u_{\rm SL} E_{\rm e,\chi}^2 \\
                                &\quad + b_{\rm Synch.}^{0}B_{\rm ISM}^2(r) E_{\rm e,\chi}^{2},
 \end{split}
 \end{equation}
where the coefficients have the values $b_{\rm Syn.}^{0}=0.0254 $ and $b_{\rm IC}^{0}=0.76 $ in units of $10^{-16} ~\rm GeV^{-1}$ \citep[e.g.,][]{Colafrancesco2006,McDaniel2018}, the photon energy densities are $u_{\rm SL}=8\, {\rm eV/cm^3}$ for starlight and $u_{\rm CMB}=0.25\,{\rm eV/cm^3}$ for CMB photons \citep[e.g.,][]{Porter2008,Profumo2010}, $E_{\rm e,\chi}=\gamma_{\rm e,\chi}m_{\rm e}c^2$ is the energy of the DMEs in units of ${\rm GeV}$, and $B_{\rm ISM}(r)$ is the strength of the interstellar magnetic field, which is assumed to be a constant, $\sim 10~ \rm \mu G$.

\begin{figure*}
\centering
\includegraphics[width=0.45\linewidth]{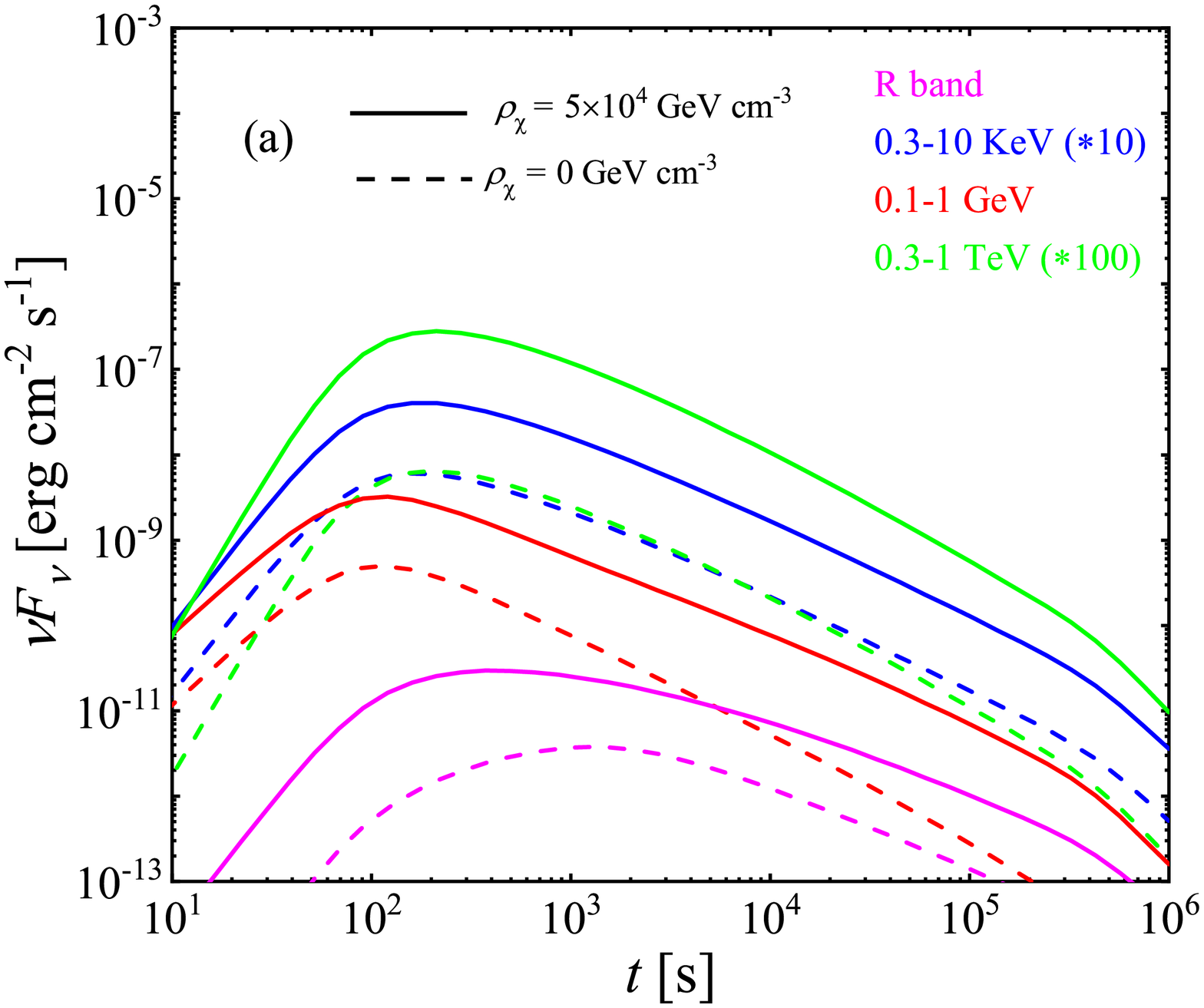}
\includegraphics[width=0.45\linewidth]{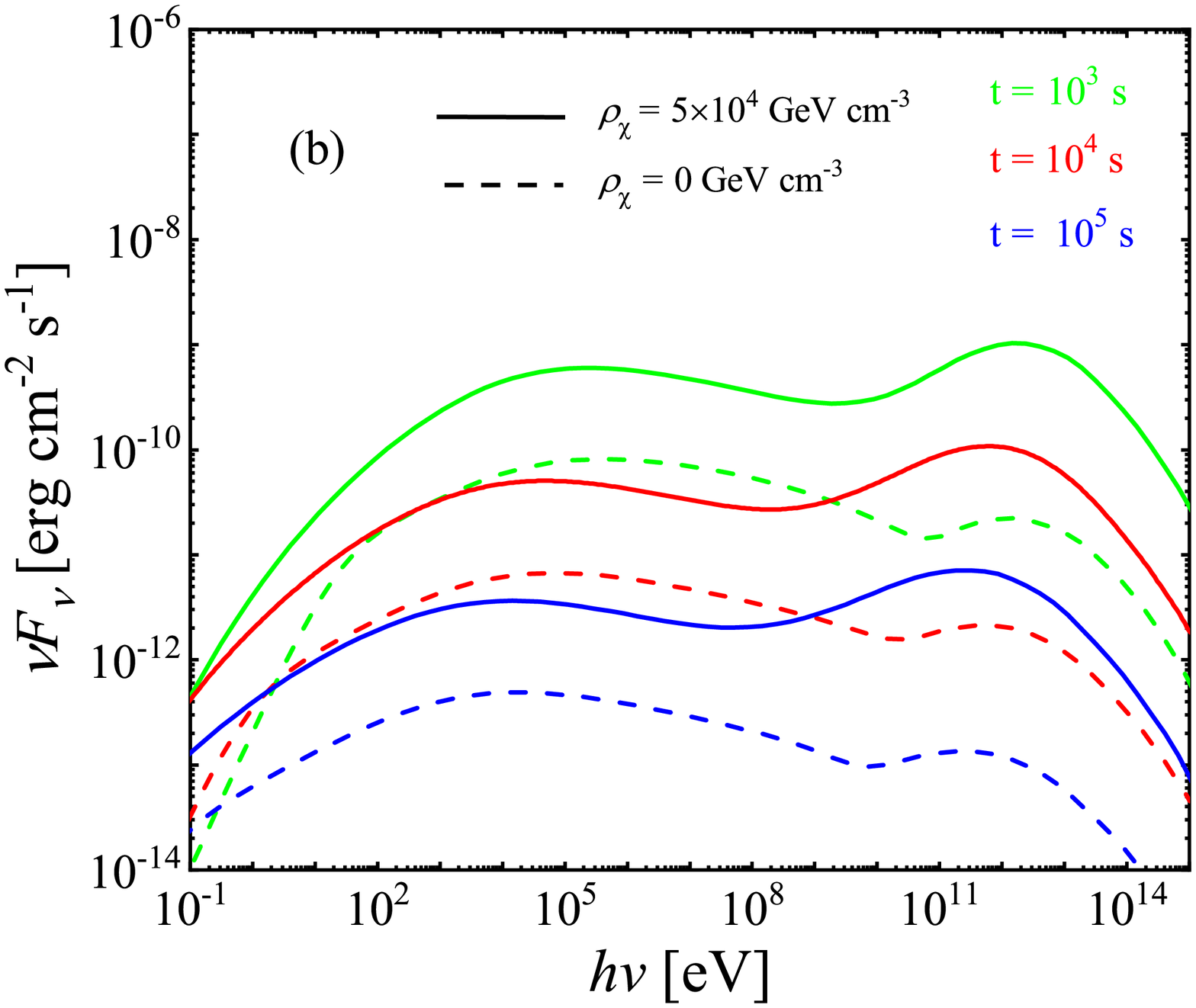}
\caption{(a) Light curves of GRB afterglows including and excluding the effects of DM with $n=0.1 \,{\rm cm^{-3}}$. DM annihilation occurs in the $b\bar{b}$ channel, and the DM particle mass is $M_{\rm \chi}=10 \, {\rm GeV}$. (b) Synchrotron $+$ SSC spectra in the same situations of (a).}
\end{figure*}

When the external forward shock encounters the circumburst medium, the DMEs are accelerated. Here, a simple scenario of Fermi-type shock acceleration is applied in which the energy of the electrons can increase by a factor on the order of $\Gamma$ during the first shock crossing and increase by a factor of $\bar{g}\sim 2.0$ in subsequent shock-crossing cycles in the frame of the shock downstream \citep[e.g.,][]{Gallant1999,Achterberg2001}. It should be noted that DMEs obtain energy of $\delta=\epsilon_{\rm e} \eta m_{\rm p}(p-2)(\Gamma-1)/m_{\rm e}(p-1)\Gamma$ in the precursor of the shock before acceleration \citep[e.g.,][]{Lemoine2017} due to the interactions of the DMEs with the electromagnetic field \citep[e.g.,][]{Kumar2015}. After every cycle, only a fraction of the DMEs can return to the upstream region, while the residual DMEs escape far downstream. According to the results of Monte Carlo simulations of test particles, we assume that the DMEs downstream have the same probability $\bar{P}_{\rm ret}(\sim0.4)$ to return to the upstream region and that the probability of returning upstream is unity \citep[e.g.,][]{Lemoine2003}. Hence, the Lorentz factors and number density spectra of the escaping DMEs after the $N$th cycle are $\gamma'^{N}_{\rm e,\chi}=\bar{g}^{N}\Gamma(\gamma_{\rm e,\chi}+\delta)$ and $f_{\rm esc}^{N}(\gamma'^{N}_{\rm e,\chi})=\bar{P}^{N}_{\rm ret}\bar{g}^{-N}(1-\bar{P}_{\rm ret}) {dn_{\rm e,\chi}(\gamma_{\rm e,\chi})}/k\Gamma{d\gamma_{\rm e,\chi}}$, respectively. Here, $k=(\gamma_{\rm e,\chi}+\delta)/\gamma_{\rm e,\chi}$. By accumulating the number density spectra of the accelerated DMEs, ${dn_{\rm e, \chi}'(\gamma_{\rm e,\chi}')}/{d\gamma_{\rm e,\chi}'}=\sum_{N=0}^{N=\infty} f_{\rm esc}^{N}(\gamma'_{\rm e,\chi})$ is obtained. The fact that this method can obtain the initial spectrum of shock-accelerated CMEs that is generally assumed to have a power-law form illustrates that simplifying the acceleration of the DMEs in this way is feasible.

It should be noted that when the number density of DMEs is much larger than that of CMEs, the value of $\eta$ is much less than unity, resulting in $\gamma^\prime_{\rm e,min}\ll \Gamma$, which conflicts with the increasing energy of DMEs during the first shock crossing. Thus, the shock-induced acceleration of DMEs in this case is inefficient. In other words, not all DMEs are efficiently accelerated. Considering the contributions of nonaccelerated DMEs simultaneously, $\hat{Q}'_{\rm DM}(\gamma_{\rm e,\chi}',R)$ can be derived as
\begin{eqnarray}
\hat{Q}'_{\rm{DM}}(\gamma_{\rm e,\chi}',R)=&&{4\pi R^2 \bigg[A\frac{dn_{\rm e,\chi}'(\gamma_{\rm e,\chi}')}{d\gamma_{\rm e,\chi}'}} \nonumber\\&&+(1-A)\frac{dn_{\rm e,\chi}(\gamma_{\rm e,\chi})}{\Gamma d\gamma_{\rm e,\chi}}\bigg],
\end{eqnarray}
where $A$ is set as $1$ when $\delta \gtrsim 1$; otherwise, the value of $A$ equals the ratio of the number density of shock-accelerated DMEs to the number density of all DMEs. Here, $\delta \simeq 1$ is treated as the boundary in the case of efficient acceleration between all and some DMEs.

\begin{figure*}
\centering
\includegraphics[width=0.45\linewidth]{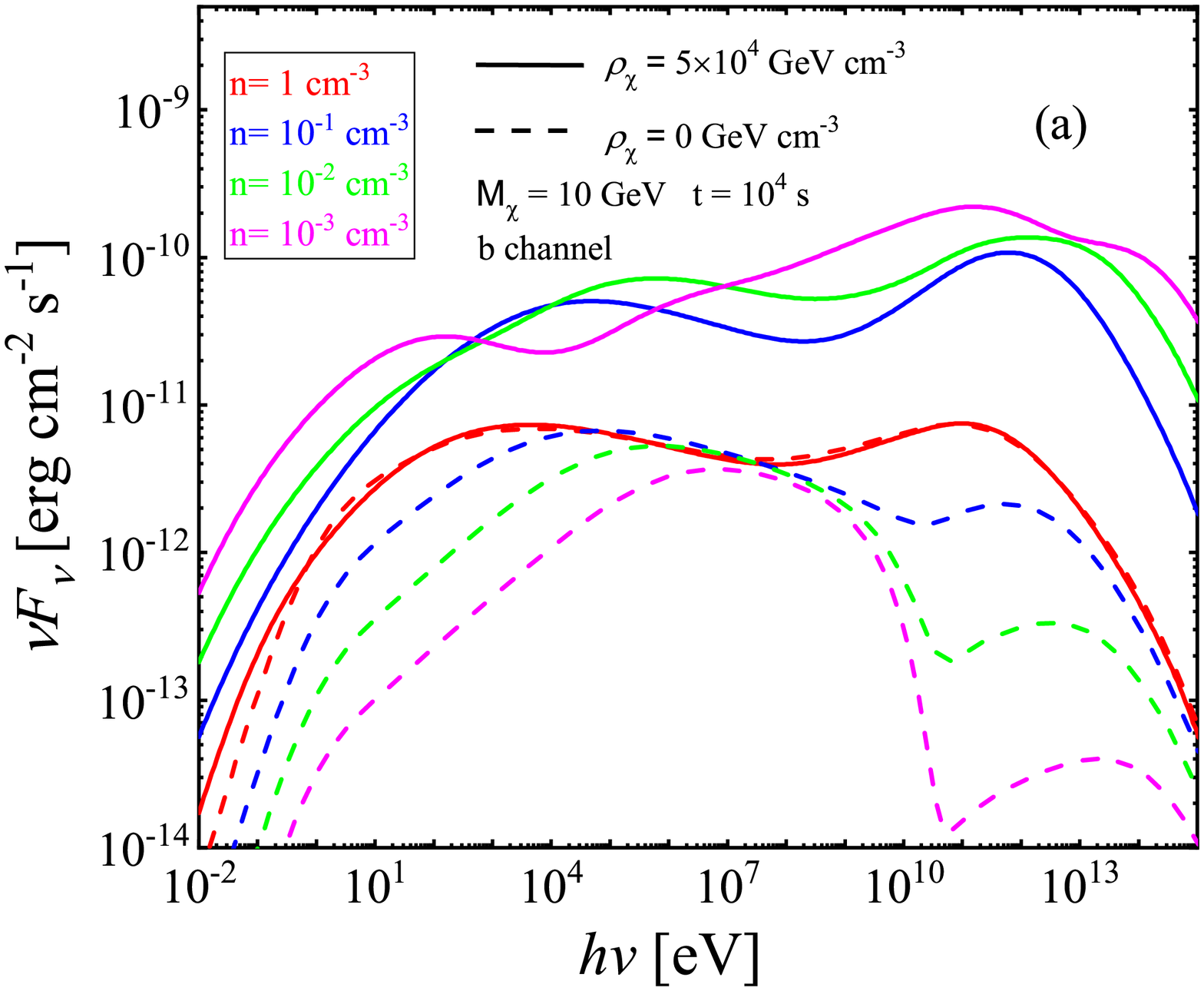}
\includegraphics[width=0.45\linewidth]{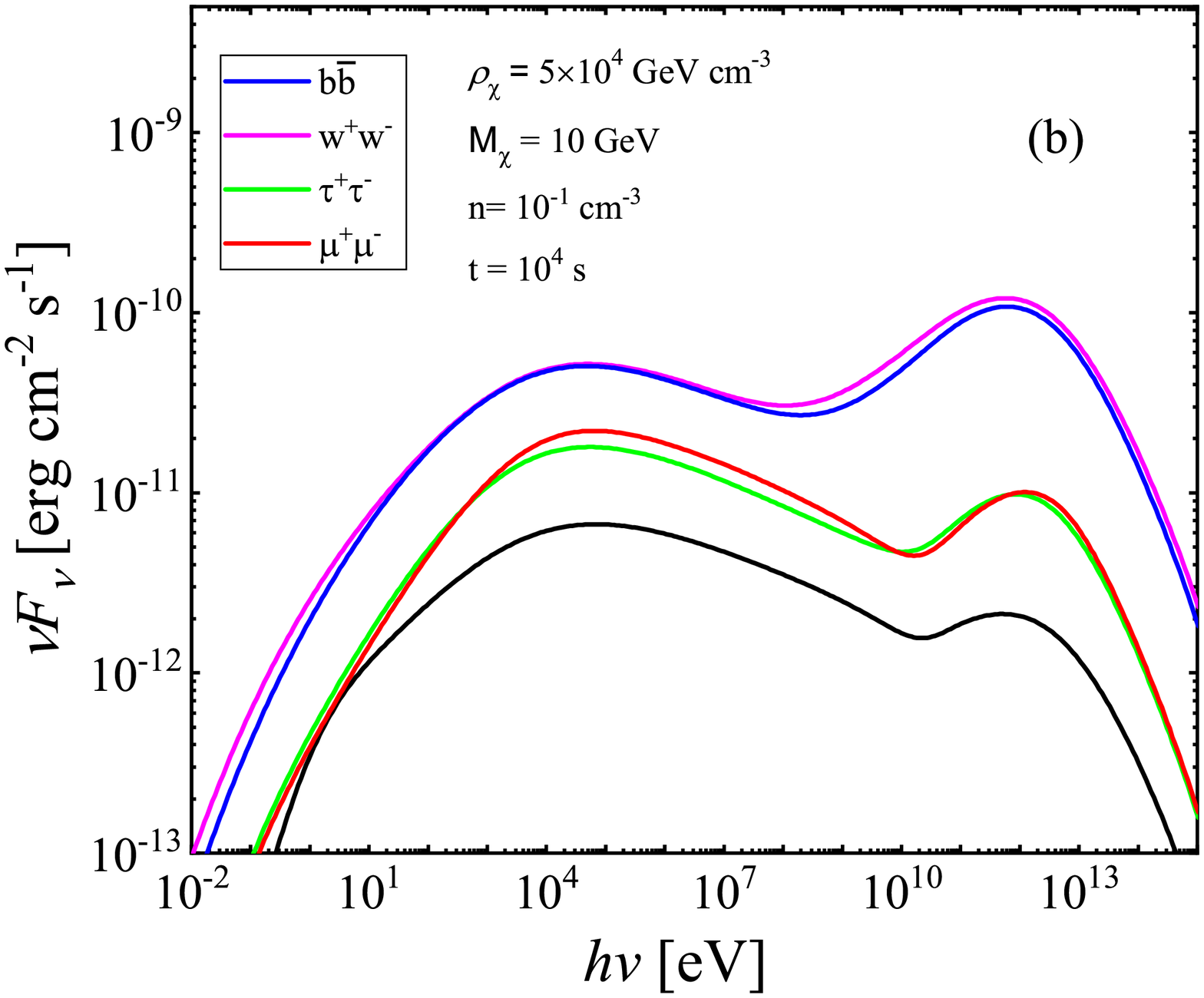}
\includegraphics[width=0.45\linewidth]{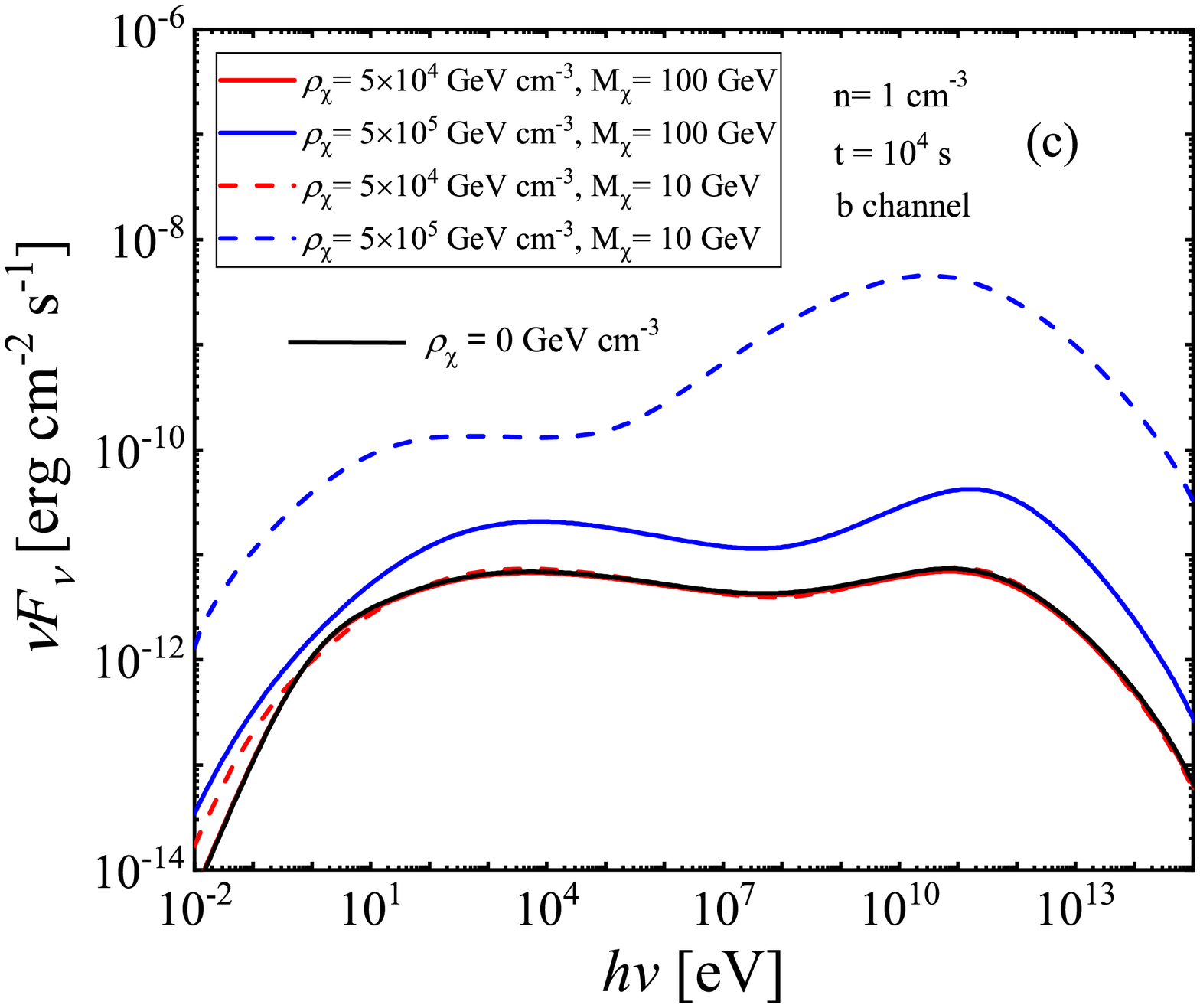}
\includegraphics[width=0.45\linewidth]{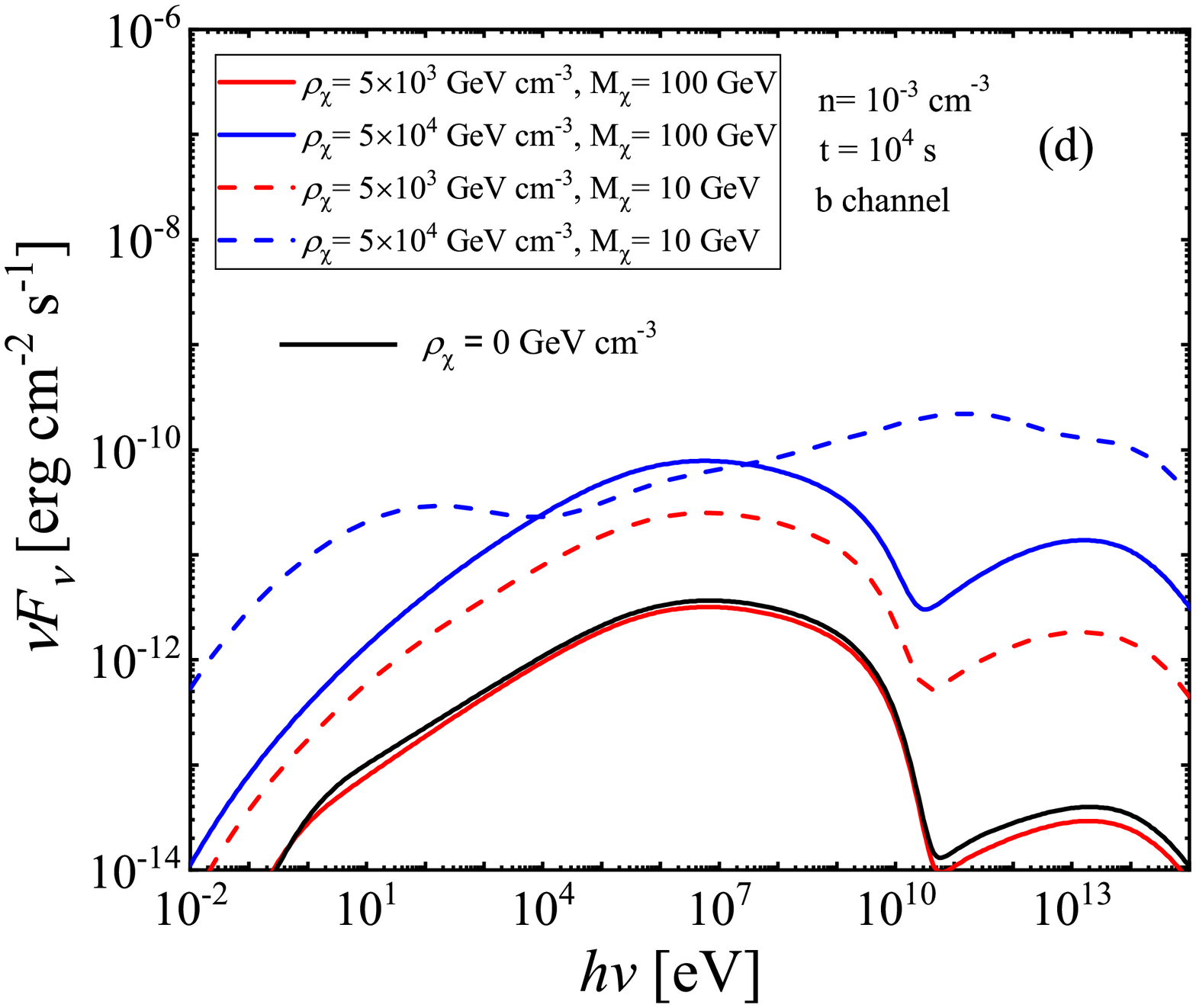}
\caption{(a) Synchrotron $+$ SSC spectra of GRB afterglows for different CME number densities ($n=1 \,{\rm cm^{-3}}$, $10^{-1} \,{\rm cm^{-3}}$, $10^{-2} \,{\rm cm^{-3}}$, and $10^{-3} \,{\rm cm^{-3}}$) at $t=10^4 \,{\rm s}$ with $\rho_{\chi}=5\times 10^4 \,{\rm GeV~cm^{-3}}$ and $M_{\rm \chi}=10 \, {\rm GeV}$ in the $b\bar{b}$ channels. The solid lines and dashed lines denote the cases including and excluding the contributions of DMEs, respectively. (b) Synchrotron $+$ SSC spectra of GRB afterglows in four different annihilation channels ($b\bar{b}$, $w^+w^-$, $\tau^+\tau^-$, and $\mu^+\mu^-$) at $t=10^4 \,{\rm s}$ with $n=10^{-1}\,{\rm cm^{-3}}$, $\rho_{\chi}=5\times 10^4 \,{\rm GeV~cm^{-3}}$, and $M_{\rm \chi}=10 \, {\rm GeV}$. The black solid line denotes the spectrum excluding the contributions of DMEs. (c) Synchrotron $+$ SSC spectra of GRB afterglows for different DM densities ($\rho_\chi$ = $5\times 10^4$ and $5 \times 10^5 \,{\rm GeV~cm^{-3}}$) and particle masses ($M_\chi$ = 10 and 100 GeV) at $t=10^4 ~{\rm s}$. DM annihilation occurs in the $b\bar{b}$ channel, and the CME number density is $n=1 \,{\rm cm^{-3}}$. The black solid line denotes the spectrum excluding the contributions of DMEs. (d) Same as (c) except $n=10^{-3} \,{\rm cm^{-3}}$ and $\rho_\chi$ = $5\times 10^3$ and $5 \times 10^4 \,{\rm GeV~cm^{-3}}$.}
\end{figure*}

\subsection{Afterglow radiation}

The synchrotron radiation power at the frequency $\nu'$ can be obtained by \citep{Rybicki1979}
\begin{equation}
P'_{\rm syn}({\nu}')=\frac{\sqrt{3} e^3 B'}{m_{\rm e}c^2}\int\nolimits_{\gamma_{\rm e,min}'}^{\gamma_{\rm e,max}'}\bigg(\frac{dN_{\rm e}'}{d\gamma_{\rm e}'} \bigg) F \bigg(\frac{\nu'}{\nu_{\rm c}'} \bigg) d\gamma_{\rm e}',
\end{equation}
where $\nu_{\rm c}'=3 e B' \gamma_{\rm e}'^{2}/4 \pi m_{\rm e}c$ and $F({\nu'}/{\nu_{\rm c}'} )=(\nu'/\nu_{\rm c}') \int\nolimits_{\nu'/\nu_{\rm c}'}^{+ \infty} K_{5/3}(x) dx$, where $K_{5/3}(x)$ is the modified Bessel function of order 5/3. The number density of SSC seed photons can be written as \citep[e.g.,][]{Fan2008}
\begin{equation}
n'_{\rm ph}(\nu') \approx \frac{1}{4\pi R^2}\frac{1}{ch\nu'} P'_{\rm syn}(\nu'),
\end{equation}
and thus, the SSC power at the frequency $\nu_{\rm ic}^{\prime}$ is \citep[e.g.,][]{Blumenthal1970}
\begin{eqnarray}
P'_{\rm SSC}(\nu'_{\rm ic})=&&\frac{3\sigma_{\rm T} c h \nu'_{\rm ic}}{4} \int\nolimits_{ \nu'_{\rm min}}^{ \nu'_{\rm max}} \frac{n'_{\rm ph}(\nu')d\nu'}{\nu'} \nonumber\\&&\times \int\nolimits_{\gamma'_{\rm e,min}}^{\gamma'_{\rm e,max}} \frac{y(q,g)}{\gamma_{\rm e}^{\prime 2}} \frac{dN'_{\rm e}}{d\gamma'_{\rm e}}d\gamma'_{\rm e},
\end{eqnarray}
where $y(q,g)=2q\,\ln q+(1+2q)(1-q)+8q^2 g^2 (1-q)/(1+4qg)$, $q=w/4g(1-w)$, $g=\gamma_{\rm e}^{\prime}h\nu^{\prime}/m_{\rm e} c^2$, and $w=h\nu_{\rm ic}^{\prime}/\gamma_{\rm e}^{\prime} m_{\rm e} c^2$. The observed spectral flux is \citep[e.g.,][]{Granot1999}
\begin{equation}
F_{\nu_{\rm obs}}=\frac{1+z}{4\pi D_{\rm L}^2}
{\int}\kern-15pt\int\limits_{(\rm EATS)}{P'(\nu '){D^3}d\Omega},
\end{equation}
where ``EATS'' is the equal-arrival time surface corresponding to the same observer time, $\nu^{\prime}=(1+z)\nu_{\rm obs}/D$ (with $D$ being the Doppler factor of the emitter), $D_{\rm L}$ is the luminosity distance in the standard $\Lambda$CDM cosmology model ($\Omega_M=0.27$, $\Omega_\Lambda=0.73$, and $H_0=71~\rm km~s^{-1}~Mpc^{-1}$), and $z$ is the redshift of the burst.

\section{Results}

Following the above method, we can calculate the light curves and spectra of GRB afterglows, including the effects of DM. The universal parameters of the external shock model are set as the isotropic kinetic energy $E_{\rm k,iso}=10^{53}\,{\rm ergs}$, $\epsilon_{\rm e}= 10^{-1}$, $\epsilon_{B}=10^{-3}$, $\Gamma = 200$, $p=2.3$, $\theta_{\rm j}=0.1$, and $z=1$.

Figure 1(a) shows the light curves of GRB afterglows including and excluding the effects of DM with $n=0.1\,{\rm cm^{-3}}$. DM annihilation occurs in the $b\bar{b}$ channel, and the DM particle mass is $M_{\rm \chi}=10 \, {\rm GeV}$. The solid and dashed lines represent the light curves and spectra with $\rho_{\chi}=5.0\times 10^4 \,{\rm GeV~cm^{-3}}$ and 0, respectively. The magenta and blue lines correspond to the synchrotron emission in the optical $R$ band and $0.3-10\,{\rm KeV}$, respectively. The red lines correspond to the synchrotron + SSC emissions in the range of $0.1-1\,{\rm GeV}$, and the green lines correspond to the SSC emissions in the range of $0.3-1 \,{\rm TeV}$. The fluxes in the ranges of $0.3-10\,{\rm KeV}$ and $0.3-1 \,{\rm TeV}$ are multiplied by $10$ and $100$, respectively. In this case, the annihilation DM effects are adequately displayed in the multiband light curves of GRB afterglows. Figure 1(b) shows the synchrotron $+$ SSC spectra in the same situation as Figure 1(a). The green, red, and blue lines represent the spectra at $10^3 \,{\rm s}$, $10^4 \,{\rm s}$ and $10^5 \,{\rm s}$, respectively. Once the effects of DM are considered, in addition to fluence increasing, the shapes of the spectra also change.

Figure 2 shows the synchrotron $+$ SSC spectra of GRB afterglows with different parameters, i.e., DM density, particle DM mass, annihilation channel, and CME density. Figure 2(a) displays the synchrotron $+$ SSC spectra of GRB afterglows with different $n$ at $t=10^4 ~{\rm s}$ after the GRB is triggered. The DM density is set to $5\times 10^4\,{\rm GeV~cm^{-3}}$ or $0\,{\rm GeV~cm^{-3}}$, and the DM particle mass is taken as 10 GeV. The red, blue, green, and magenta lines represent $n=1.0\,{\rm cm^{-3}}$, $10^{-1}\,{\rm cm^{-3}}$, $10^{-2}\,{\rm cm^{-3}}$, and $10^{-3}\,{\rm cm^{-3}}$, respectively. The solid and dashed lines represent the cases including and excluding the contribution of the DMEs, respectively. Different CME densities result in different spectral shapes, and the inclusion of DMEs further makes the variety of the spectra. For a lower CME density, the DME effect is more significant.

The spectra in four different annihilation channels are presented in Figure 2(b). It is obvious that the fluxes in the $w^+w^-$ and $b\bar{b}$ annihilation channels are higher than those in the $\tau^+\tau^-$ and $\mu^+\mu^-$ annihilation channels.

Figure 2(c) shows the synchrotron $+$ SSC spectra of GRB afterglows for different DM densities and DM masses at $t=10^4 ~{\rm s}$. The values of the DM density is set to $5\times 10^4\,{\rm GeV~cm^{-3}}$ and $5 \times 10^5 \,{\rm GeV~cm^{-3}}$, and the values of DM particle mass are taken as 10 GeV and 100 GeV. The DM annihilation channel is assumed to be the $b\bar{b}$ channel, and the CME number density is $n=1 \,{\rm cm^{-3}}$. Figure 2(d) presents the afterglow spectra for the same parameters of Figure 2(c), except the parameters $n=10^{-3} \,{\rm cm^{-3}}$, and $\rho_\chi$ = $5\times 10^3\,{\rm GeV~cm^{-3}}$ and $5 \times 10^4 \,{\rm GeV~cm^{-3}}$. The black solid lines in Figures 2(c) and 2(d) denote the spectra excluding the contributions of the DMEs. The flux values of the black lines are slightly higher than those of the red solid lines (corresponding to lower DM densities and high DM particle masses) because $\gamma_{\rm e,min}$ decreases slightly when the effects of DM are considered. According to all the lines, a higher DM density and lower DM particle mass can dramatically increase the fluxes. As shown by the blue dashed line in Figure 2(d), the flux exhibits prominent improvements in the low- and high-energy parts due to the contributions of inefficiently accelerated DMEs and the increased number of seed photons, respectively.

Since many parameters related to jets are uncertain and the ISM and DM can prominently influence the light curves and spectra of GRB afterglows, it is difficult to distinguish the distributions of DM unless the properties of the circumburst medium can be constrained by observations. For a full description of the medium properties of GRB afterglows, the location of GRBs in the galaxy should be considered, as discussed below.

\section{GRB Location in the Galaxy}

To simplify, we assume a DM halo with the virial mass $M_{\rm vir}$ in the virial radius $R_{\rm vir}$. The mean density is equal to the virial overdensity $\Delta_{\rm vir}$ times the mean background density $\rho_{\rm u}$, given by \citep[e.g.,][]{Bullock2001}
\begin{equation}
M_{\rm vir}\equiv\dfrac{4\pi}{3}\Delta_{\rm vir} \rho_{\rm u} R_{\rm vir}^{3}.
\end{equation}
$\Delta_{\rm vir}=200$ is generally preferred, which is independent of cosmology. Thus, the corresponding virial radius and virial mass are $R_{\rm 200}$ and $M_{\rm 200}$, respectively. Furthermore, the concentration parameter \citep[e.g.,][]{Bullock2001} can be defined as
\begin{equation}
c_{200} \equiv \dfrac{R_{200}}{r_{\rm c}},
\end{equation}
which is also related to the virial mass \citep[e.g.,][]{Buote2007}, i.e.,
\begin{equation}
c_{200}=\frac{6.9}{(1+z)}\bigg(\frac{M_{\rm 200}}{10^{14}~M_\odot}\bigg)^{-0.178}.
\end{equation}
Moreover, the virial mass can be derived by integrating different DM density profiles in the viral radius $R_{200}$,
\begin{equation}
M_{200} = 4\pi \int_{0}^{R_{200}} \rho(r)r^{2}\,dr.
\end{equation}

\begin{figure}[t]
\includegraphics[width=0.95\linewidth]{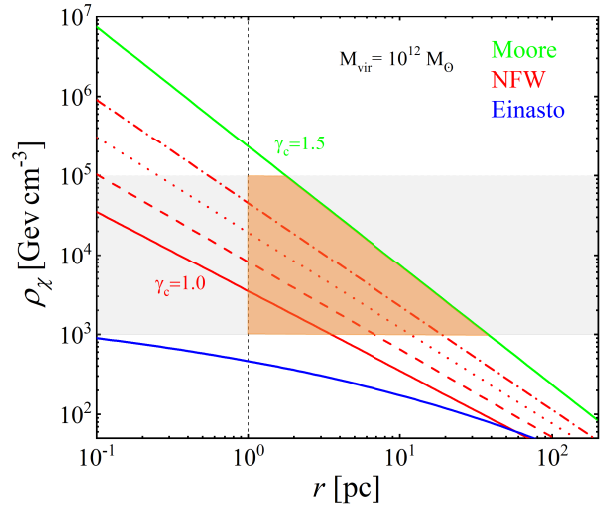}
\caption{DM density as a function of the distance to the center of a massive galaxy with the virial mass $\sim 10^{12}\, {M}_{\odot}$. The lines of different colors denote different DM density profiles. In addition, the red solid, dashed, dashed-dotted, and dotted lines denote the results with the following values of the inner slope of the NFW DM density profile: $1.0$, $1.1$, $1.2$, and $1.3$, respectively.}
\end{figure}

The DM density of the galaxies is a function of radius. The general Navarro-Frenk-White (NFW) density profile reads \citep{Navarro1997}
\begin{eqnarray}
\rho_{\rm NFW}(r)=\frac{\rho_{\rm c}}{(r/r_{\rm c})^{\gamma_{\rm c}}[1+(r/r_{\rm c})]^{3-{\gamma_{\rm c}}}},
\end{eqnarray}
where $\rho_{\rm c}$ and $r_{\rm c}$ are the characteristic density and radius, respectively. Here, we consider four values of the slope: $\gamma_{\rm c}=1.0$, $1.1$, $1.2$, and $1.3$, respectively. The profile proposed by \cite{Moore1999} corresponding to the slope $\gamma_{\rm c}=1.5$ is derived according to N-body simulations, which suggests a steeply rising DM density towards the galaxy center. In addition, the Einasto law is written as \citep{Merritt2006}
\begin{eqnarray}
\rho_{\rm Einasto}(r)=\rho_{\rm c} \exp[-d_{N}(({r}/{r_{\rm c}})^{1/N}-1)],
\end{eqnarray}
where $d_N \approx 3N-1/3+0.0079/N$ for $N \gtrsim 0.5$.

Combining the above equations, one can obtain the characteristic radius and density of the DM density profiles with the virial mass $M_{\rm 200}= 10^{12}\, {M}_{\odot}$. The results are shown in Figure 3. The blue and green solid lines represent the Einasto law and Moore law, respectively. The red lines represent the NFW formula, where the solid, dashed, dotted, and dashed-dotted lines denote the NFW profile with the following four values of the inner slope: $1.0$, $1.1$, $1.2$, and $1.3$, respectively. The gray shaded region indicates the DM density range of $\sim10^3 \,- \,10^5 \,{\rm GeV~cm^{-3}}$. The vertical black dashed line denotes the location 1 pc from the center of the galaxy. According to the magnetic field model used in the galaxy center \citep{Aloisio2004}, $B_{\rm ISM}$ increases rapidly with decreasing radius within $1\,{\rm pc}$, so the value of $B_{\rm ISM}$ assumed above would be inappropriate within $1\,{\rm pc}$. Finally, the brown shaded region represents the possible locations of GRBs with significant contributions from DM, i.e., several to tens of parsecs from the center of a galaxy with virial mass $\sim 10^{12}\, {M}_{\odot}$.

Moreover, the effects of DM spatial diffusion \citep[e.g.,][]{Colafrancesco2006}, re-acceleration, and advection \citep[e.g.,][]{Strong2007} might be important at the galaxy scale or beyond, which would decrease the DME densities around GRBs. These effects are neglected for our calculations.

\section{Conclusions and discussion}

In this paper, we study the contributions of DM annihilation to GRB afterglows. If GRBs occur at distances of several to tens of parsecs from the centers of massive galaxies, the effects of DMEs should be significant. Moreover, the influences of different DM particle masses and annihilation channels on GRB afterglows are adequately reflected in the flux changes.

To date, no GRB has been discovered at a distance of tens of parsecs from a galaxy center, but one may expect that these events could be detected in the future because the density of stars near the galaxy center is generally higher than that in other regions within the host galaxy. Observations show that the majority of stars formed at distances of a few parsecs from the center of the Milky Way are older than $5$ Gyr \citep[e.g.,][]{Blum2003}, and one pulsar, PSR J$1745-2900$, was detected $\sim$ 0.12 pc from Sagittarius $\rm A^\star$ \citep{Eatough2013}. Of course, the dense gas and bright galaxy center will trouble the observations of those distant GRBs.

In addition to DMEs, gamma-ray emission is another final product of DM annihilation due to $\pi^0$ decay. The gamma-ray flux produced by $\pi^0$ decay from a galaxy with $M_{\rm vir}= 10^{12}\, M_{\odot}$ at $z=1$ is far below the observational threshold and hence can be neglected. However, if this galaxy is located near the Milky Way, the gamma-ray flux can be detected. Therefore, we use observations of a neighboring galaxy, M31, which has a virial mass of $\sim 10^{12}\,M_{\odot}$, to test our model under the assumption that the gamma-ray emission near the center of M31 originated from DM annihilation. In the observation, a gamma-ray excess with a flux of $(5.6\pm0.6) \times10^{-12} \,{\rm erg\, cm^{-2}\,s^{-1}}$ was detected in the energy range from $0.1\,{\rm GeV}$ to $100\,{\rm GeV}$ within the central M31 region of $\sim 5\,{\rm kpc}$ \citep{Ackermann2017}. We can calculate the gamma-ray flux by the formula
\begin{eqnarray}
\frac{d\Phi_{\gamma}(E_{\gamma})}{dE_{\gamma}}=&&\frac{1}{ D_A^2}\frac{\langle\sigma_{A} \upsilon\rangle}{2m_{\chi}^2} \frac{dN_{\gamma}(E_{\gamma})} {dE_{\gamma}}E_{\gamma}^2 \nonumber \\&& \times \int\nolimits_{r_{\rm min}}^{r_{\rm max}} \rho_{\chi}^2(r) r^2 dr,
\end{eqnarray}
where ${dN_{\gamma}(E_{\gamma})}/{dE_{\gamma}}$ is the gamma-ray spectrum corresponding to $M_{\chi}=100\,{\rm GeV}$ and the $b\bar{b}$ annihilation channel and $r_{\rm max}=5\,{\rm kpc}$, and $D_A$ is the distance of M31. The NFW profile with $\gamma_{\rm c}=1.3$ is adopted. By integrating the energy range from $0.1\,{\rm GeV}$ to $100\,{\rm GeV}$, the gamma-ray flux is $5.7 \times 10^{-13} \,{\rm erg\, cm^{-2}\,s^{-1}}$. This flux is less than that observed in M31, indicating that our model is self-consistent.

\acknowledgments
This work was supported by the National Natural Science Foundation of China under grants 11822304, 11890692, 11773007, and U1531130, and the Guangxi Science Foundation under grant 2018GXNSFFA281010.

\end{document}